\documentclass[floatfix,pre,twocolumn,showpacs]{revtex4}

\newcommand{\CBPF}{$^1$Centro Brasileiro de Pesquisas Fisicas and National Institute of Science and Technology for Complex Systems - Rua Xavier Sigaud
150, 22290-180 Rio de Janeiro-RJ, Brazil} 
\newcommand{\SFI}{$^2$Santa Fe Institute, 1399 Hyde Park Road, Santa Fe, NM 87501, USA}
\usepackage[]{graphicx}
\usepackage{bm}
\usepackage{amsmath}

\begin{document}

\title{Time evolution towards $q$-Gaussian stationary states through unified It\^o-Stratonovich stochastic equation}
\author{B. Coutinho dos Santos$^1$}
\email{bernardoc@cbpf.br}
\author{C. Tsallis$^{1,2}$}
\email{tsallis@cbpf.br}
\affiliation{\CBPF \\ \SFI}
\date{\today} 
\begin{abstract}
We consider a class of single-particle one-dimensional stochastic equations
which include external field, additive and multiplicative noises. We use a
parameter $\theta \in [0,1]$ which enables the unification of the traditional
It\^o and Stratonovich approaches, now recovered respectively as the $\theta=0$
and $\theta=1/2$ particular cases to derive the associated Fokker-Planck equation
(FPE). These FPE is a {\it linear} one, and its stationary state is given by a
$q$-Gaussian distribution with $q = \frac{\tau + 2M (2 - \theta)}{\tau + 2M (1 - \theta)}<3$,
where $\tau \ge 0$ characterizes the strength of the confining external field, and $M \ge 0$
is the (normalized) amplitude of the multiplicative noise.
We also calculate the standard kurtosis $\kappa_1$ and the $q$-generalized kurtosis  $\kappa_q$
(i.e., the standard kurtosis but using the escort distribution instead of the direct one). Through
these two quantities we numerically follow the time evolution of the distributions. Finally, we
exhibit how these quantities can be used as convenient calibrations for determining the index
$q$ from numerical data obtained through experiments, observations or numerical computations.
\end{abstract}
\pacs{02.50.-r, 05.20.-y, 05.40.-a, 05.90.+m}

\maketitle

\section{Introduction}

The random walk is the simplest model of diffusive processes in
physics. If there is no wind, nor any other source of symmetry-breaking, a drunk has probability 1/2 to take a step to
the right, and the same probability to take a step to the left, at each
instant. Underneath this model there is a formal mathematical theory known as
stochastic calculus, which in some sense can be interpreted as a extension of
the standard differential calculus taught in undergraduate courses. The
stochastic as well as the standard calculus are based on the definition of
the integral. Let us define the stochastic integral by
\begin{equation}
 I[G(t)] = \int_{t_0}^{t} dW(t^{\prime}) \, G(t^{\prime}) \; ,
\end{equation}
where $G(t)$ is a left-continuous function (i.e., a function which is continuous from the left at all the points where it is defined) and $W(t)$ is a Wiener process
\cite{Gardiner,Risken}. As in the Riemann integral definition, the formal stochastic
integral (also known as Riemann-Stieltjes integral) is a infinite discrete
sum of very small intervals ($dW(t^{\prime})$) of a stochastic function. When we perform this sum
we must make a choice, more precisely the function $G(t^{\prime})$ has to be evaluated in some
point inside each interval. This choice defines what kind of stochastic calculus
will be performed henceforth. The two most famous procedures are It\^o calculus
and Stratonovich calculus. In the first, It\^o calculus, the function is
evaluated at the beginning of the intervals:
\begin{equation}
 I_I[G(t)] = \text{ms -}\hspace{-0.1cm}\lim_{n \rightarrow \infty} \sum_{i=1}^{n} G_{i-1}\Delta W_i \; ,
\end{equation}
where ($\text{ms -}\lim$) stands for {\it mean squared} limit, i.e. a second
moment convergence; $n$ is the number of subintervals in which we divide
the interval $[t_0,t]$, and $\Delta W_i = W_{i} - W_{i-1}$. In the Stratonovich calculus we take the arithmetic average
between the integrate function values at the beginning and at the end of the
intervals, as follows
\begin{equation}
 I_S[G(t)] = \text{ms -}\hspace{-0.1cm}\lim_{n \rightarrow \infty} \sum_{i=1}^{n}
\dfrac{\left[G_{i} + G_{i-1}\right]}{2} \Delta W_i \; .
\end{equation}

An unified form can be proposed \cite{Germano,Lau,Lancon1, Lancon2} for the
stochastic integral, namely
\begin{equation}\label{proposal}
 I_\theta[G(t)] = \text{ms -}\hspace{-0.1cm}\lim_{n \rightarrow \infty} \sum_{i=1}^{n}
\left[\theta G_{i} + (1-\theta) G_{i-1}\right] \Delta W_i \; ,
\end{equation}
where $0 \le \theta \le 1$. Notice that the two traditional procedures can
be recovered easily. Indeed, if $\theta=0$ we obtain the It\^o approach,  and
if  $\theta=1/2$ we obtain the Stratonovich approach. Moreover, if $\theta=1$
we recover the so called backward-It\^o \cite{backward} stochastic approach,
that also known as isothermal convention \cite{Lau,Lancon1,Lancon2}.

It is possible to go one step further and generalize (\ref{proposal}). Indeed, we can assume that the values of $\theta$, at each interval $dW(t^{\prime})$, are given by an arbitrary distribution $\rho(\theta)$ ($\int_0^1 d\theta \,\rho(\theta)=1$). In particular, if this distribution is $\rho(\theta)=\delta(\theta-\theta_0)$, we recover the three cases mentioned above for suitable values of $\theta_0$, namely $\theta_0=0$ (It\^o), $\theta_0=1/2$ (Stratonovich) and $\theta_0=1$ (backward-It\^o). An interesting remark arises when $\rho(\theta)$ is constant. In this case $\left\langle \theta\right\rangle=1/2$ which coincides precisely with the average value corresponding to the (frequently adopted) Stratonovich approach. In some sense, this is what seems to happen in most experiments, where the act of measuring is not instantaneous but in a time window.
Let us finally mention that, in some sense, the only hypothesis strictly consistent with causality \cite{causality} is $\rho(\theta)=\delta(\theta)$.

An instructive example of the aplication of definition (\ref{proposal}) consists in the
choice $G(t)=W(t)$, so
\begin{widetext}
\begin{eqnarray}\label{example}
\nonumber
 I_\theta[W(t)] \equiv\int_{t_0}^{t} dW(t^{\prime}) W(t^{\prime}) &=& \text{ms -}\hspace{-0.1cm}\lim_{n \rightarrow
\infty} \sum_{i=1}^{n} \left[\theta W_{i} + (1-\theta)
W_{i-1}\right] \Delta W_i \\
\label{unification}&=& \frac{1}{2} \left\{ [W(t)]^2  - [W(t_0)]^2 - (1 -
2\theta) (t - t_0)
\right\}\; .
\end{eqnarray}
\end{widetext}
Again, when $\theta=0$ ($\theta=1/2$) we obtain the standard result for
the It\^o (Stratonovich) approach. For more details about the derivation
of the last equation, see Appendix.

In addition to the ingredients of stochastic calculus that we have mentioned above, some other concepts will be necessary in the present paper. In particular, 
escort mean values \cite{Beckbook,Tsallis1998} are convenient theoretical tools for describing basic features of
some probability densities, mainly those densities which decay as power laws
that naturally appear in the study of complex systems dynamics, such as those
obeying nonextensive statistical mechanics \cite{Tsallis2009}. This theory generalizes the Boltzmann-Gibbs (BG) statistical mechanics, and is governed  by an entropic index $q$, which equals unity for the BG case. The characterization of a
probability density by its set of escort mean values, if all of them converge,
is a natural extension of the well-known characterization of a distribution in
terms of its standard moments, which corresponds to $q=1$. The $q$-generalized theory has been applied to calculate many features of several complex systems, such as: (i) The velocity distribution of cells of {\it Hydra viridissima}, that follows a 
$q=3/2$ PDF \cite{UpadhyayaRieuGlazierSawada2001}; (ii) The velocity distribution of (cells of) 
{\it Dictyostelium discoideum}, that follows a $q=5/3$ PDF in the vegetative state and a $q=2$ PDF in the starved 
state \cite{Reynolds2010}; (iii) The velocity distribution in defect turbulence \cite{DanielsBeckBodenschatz2004}; (iv)
The velocity distribution of cold atoms in a dissipative optical lattice 
\cite{DouglasBergaminiRenzoni2006}; (v) Velocity distribution during silo drainage \cite{ArevaloGarcimartinMaza2007a,ArevaloGarcimartinMaza2007b}; 
(vi) The velocity distribution in a driven-dissipative 2D dusty plasma, 
with $q=1.08\pm0.01$ and $q=1.05\pm 0.01$ at temperatures of $30000 \,K$ and $61000\, K$ respectively 
\cite{LiuGoree2008}; (vii) The spatial (Monte Carlo) distributions of a trapped $^{136}Ba^+$ ion cooled by various 
classical buffer gases at $300\,K$ \cite{DeVoe2009}; (viii) The distributions of price returns at the stock exchange 
\cite{Borland2002a,Borland2002b,Queiros2005}; (ix) The distributions of returns of magnetic field fluctuations 
in the solar wind plasma as observed in data from  Voyager 1 \cite{BurlagaVinas2005} and from Voyager 2 
\cite{BurlagaNess2009}; (x) The distributions of returns of the avalanche sizes in the Ehrenfest's dog-flea 
model \cite{BakarTirnakli2009}; (xi) The distributions of returns of the avalanche sizes in the self-organized 
critical Olami-Feder-Christensen model, as well as in real earthquakes 
\cite{CarusoPluchinoLatoraVinciguerraRapisarda2007}; (xii) The distributions of angles in the $HMF$ model 
\cite{MoyanoAnteneodo2006}; (xiii) The distribution of stellar rotational velocities in the Pleiades 
\cite{CarvalhoSilvaNascimentoMedeiros2008}; (xiv) The distribution of transverse
momenta in high-energy proton-proton collisions \cite{CMS2010}; (xv) In the ation of spin glasses
\cite{PickupCywinskiPappasFaragoFouquet2009}.

In the present work we use Eq. (\ref{unification}) to write unified Langevin
and Fokker-Planck equations. In section \ref{FPE}, after deriving these equations, 
we will show that the steady solutions of the Fokker-Planck equation (FPE) are
$q$-Gaussians (see later on for their precise definition) whose entropic parameter $q$ depends on the noise and drift
amplitudes, and also on the choice of a stochastic approach represented by the parameter
$\theta$. In section \ref{escort} we introduce a generalized kurtosis based
on \cite{Tsallis1998} to characterize the temporal
evolution of the distribution. After that, we integrate numerically the FPE with
an initial distribution different from its asymptotic form. In particular, we
consider as initial distributions $q$-Gaussians  characterized by an index $q_i$ ($q_i\neq q$). Our numerical results show how the convergence
towards the attractor behaves as a function of the parameter $\theta$. Finally,
we compare the standard kurtosis ($q=1$) with the one calculated using escort
mean values, namely $q$-kurtosis. Our results show that the standard kurtosis
has a divergence at $q=7/5$ while the $q$-kurtosis has no divergence in
the range $-1 < q \leq 3$. In addition to that, we show that the standard kurtosis and the $q$-kurtosis are monotonic
functions of the entropic parameter $q$, which suggests that they could be used as
calibration curves to determine, from numerical data, the most appropriate value of $q$.
Let us now describe the consequences of the unification
on the Langevin and Fokker-Planck descriptions.

\section{Fokker-Planck equation and its solutions}\label{FPE}

A stochastic differential equation (SDE) is not completely defined by itself. If the Langevin equation has multiplicative noise, we must choose of what approach, It\^o or Stratonovich, will be used to integrate it. This is the well-known 
It\^o-Stratonovich dilemma \cite{Gardiner,Risken, Kampen} and is ultimately solved by
taking into account the specific features of the system under investigation. 
For instance, if the noise has a finite correlation time $\tau_c$
(even if the limit $\tau_c\rightarrow0$ is used to derive the SDE) or the noise comes from external sources \footnote{External sources are those which have a parameter which permits one, in principle, to turn off the noise and those which is not influenced by the system itself. In contrast, internal noise sources are those which fluctuations are due to the fact that system itself consists of discrete particles, they are an inherent part of the mechanism by which the state of the system evolves and cannot be turned off by manipulating a parameter \cite{van_KampenJSP}.}, the Stratonovich choice is the adequate one. Instead, the It\^o formalism is the correct choice if $\tau_c$ is strictly zero or the noise comes from internal sources \cite{Gardiner,Kampen}. A very interesting and enlightening discussion about the Stratonovich-It\^o dilemma can be found in \cite{van_KampenJSP}.
This controversy between the two approaches appears in the form of the so called noise-induced drift which
is an effect of the state dependence of the noise strength.

Based on the proposal (\ref{proposal}) we will write an unified Fokker-Planck equation that contains,
as special cases, the It\^o ($\theta=0$) and Stratonovich ($\theta=1/2$) forms. To do it, let us consider the quite general 
SDE for a stochastic variable $u(t)$:
\begin{equation}\label{langevin_general}
du(t) = a[u(t),t] \; dt + b[u(t),t] dW(t)\; ,
\end{equation}
where $a[u(t),t]$ is a deterministic external force given, for instance, by a
potential, $b[u(t),t]$ is the state-dependent noise amplitude, and $dW(t) = \xi(t) dt$ is a Wiener increment ($\xi(t)$ is a Gaussian, zero mean, white noise). We will use a shorter notation for the dynamic variable $u(t)$ omitting its time dependence, so from now on $u \equiv u(t)$, $a[u(t),t] \equiv a(u,t)$ and $b[u(t),t] \equiv b(u,t)$, unless it is indispensable for the clarity of the text.

We are interested in the dynamics of the distribution function of $u$, so once we have the Langevin equation we may obtain a FPE for the probability density $P(u,t)$ by the Kramers-Moyal
expansion \cite{Risken} $\partial_t P = \sum_{n \geq 1} (-\partial_u)^n [D^{(n)}P]$,
where the coefficients $D^{(n)}$ are given by
%
\begin{equation} \label{coef}
D^{(n)}(x,t) = \frac{1}{n!}\lim_{\epsilon \rightarrow 0} \frac{\langle[u(t+\epsilon)
- x]^n\rangle}{\epsilon}\rvert_{u(t)=x} \; .
\end{equation}
%
To calculate these coefficients, we need to write the Langevin equation in the integral form
\begin{equation}\label{langevin_integral}
u(t + \epsilon) - x = \int_{t}^{t + \epsilon} dt' a[u(t'),t'] \; + \int_{t}^{t + \epsilon} dW(t') \;b[u(t'),t'] \,,
\end{equation}
and assume that $a(u,t)$ and $b(u,t)$ can be expanded as
\begin{eqnarray}
\label{coef1} a[u(t + \epsilon),t + \epsilon] &=& a[x,t + \epsilon] + a'[x,t + \epsilon] \epsilon + ... \\
\label{coef2} b[u(t + \epsilon),t + \epsilon] &=& b[x,t + \epsilon] + b'[x,t + \epsilon] \epsilon + ... \, ,
\end{eqnarray}
where $a'$ and $b'$ means differentiation with respect to $u$. Putting (\ref{coef1}) and (\ref{coef2}), up to first order on $\epsilon$ \footnote{Higher orders in $\epsilon$ will vanishes when we take the limit $\epsilon \rightarrow 0$ in the limit (\ref{coef}).}, into (\ref{langevin_integral}) and iterating the result, we get
\begin{widetext}
\begin{eqnarray}
\left\langle u(t + \epsilon) - x\right\rangle  &=& \epsilon a(u,t + \epsilon) +
b^{\prime}(u,t + \epsilon) b(u,t + \epsilon)
\left\langle \int_t^{t + \epsilon} W(t^{\prime}) dW(t^{\prime}) \right\rangle \\
&=& \epsilon \left[ a(u,t + \epsilon) + \theta
b^{\prime}(u,t + \epsilon) b(u,t + \epsilon) \right]
\; .
\end{eqnarray}
\end{widetext}
If the noise in the Langevin equation is $\delta$-correlated, after repeated iterations it is possible to show that, at $\epsilon$-order, the terms of the Kramers-Moyal expansion with $n \geq 3$ vanish. Using this arguments we can calculate the first two coefficients of that expansion, namely
\begin{eqnarray}
 K_\theta(u,t) \equiv D^{(1)}(u,t) &=& a(u,t) + \theta b^{\prime}(u,t) b(u,t) \\
 D(u,t) \equiv D^{(2)}(u,t) &=& b^2(u,t) \; .
\end{eqnarray}
This procedure leads to the following Fokker-Planck equation:
%
\begin{eqnarray}\label{FPE1}
 \frac{\partial P}{\partial t} = -\frac{\partial}{\partial u} \left\lbrace K_\theta(u,t) P - \frac{1}{2}\frac{\partial}{\partial u} \left[D(u,t) P \right] \right\rbrace \; .
\end{eqnarray}
%
We recover, for $\theta = 0$ and $\theta = 1/2$, respectively, the It\^o and Stratonovich forms currently found in the literature.
As expected, the difference between the two forms appears in the drift term $K_\theta$, not in the diffusion term. If we deal with a FPE in the Stratonovich form we should include the above mentioned noise-induced drift term $\frac{1}{2}\partial_u [D(u)]$, which is unnecessary in the It\^o form.
We can write these FPE as a continuity-like equation $\partial_t P = -\partial_u
j(u)$, if we define $j(u) = K_\theta (u)P - \frac{1}{2}\partial_u [D(u) P]$ as a probability
current. In this  case, the stationary solution of the FPE can be obtained elegantly from zero flux
boundary conditions $j(\pm \infty)=j(u)=0$.
It is very interesting to notice that this probability current can be rewritten as $j(u) = K_0 (u)P - \frac{1}{2}[D(u)]^{2\theta}\partial_u \left\lbrace [D(u)]^{1-2\theta} P\right\rbrace$, which allow us to rewrite the FPE as:
\begin{equation}\label{FPE2}
 \frac{\partial P}{\partial t} = -\frac{\partial}{\partial u} \left\lbrace K_0 (u)P - \frac{1}{2}[D(u)]^{2\theta}\frac{\partial}{\partial u} \left\lbrace [D(u)]^{1-2\theta} P\right\rbrace \right\rbrace \; .
\end{equation}

Notice that here, in contrast with Eq. (\ref{FPE1}), $\theta$ only appears in the diffusion term, and not in the drift one. The demonstration of the equivalence between Eqs. (\ref{FPE1}) and (\ref{FPE2}) is straightforward and can be found in the Appendix \ref{apendice2}.

Let us now consider a family of models represented by Langevin equations
of the type:
\begin{equation}\label{langevin1}
\dot{u} = f(u) + g(u) \xi(t) + \eta(t) \; ,
\end{equation}
where $\xi(t)$ and $\eta(t)$ are uncorrelated zero-mean Gaussian white noises
with autocorrelation function given by:
\begin{equation}\label{autocorr}
\langle \xi(t) \xi(t^{\prime}) \rangle = 2 M \delta(t - t^{\prime}) \;, \quad
\langle \eta(t) \eta(t^{\prime}) \rangle = 2 A \delta(t - t^{\prime}) \; ,
\end{equation}
where $M\geq0$ and $A>0$. The Eq. (\ref{langevin1}) can be rewritten as:
\begin{equation}\label{langevin2}
\dot{u} = f(u) + \tilde{g}(u) \zeta(t) \; ,
\end{equation}
where the additive and multiplicative noise terms were replaced by an effective
multiplicative noise given 
by $\tilde{g}(u) = \sqrt{(M [g(u)]^2 + A)/C}$ and $\zeta(t)$ being a zero-mean
Gaussian white noise with 
autocorrelation function given by $\langle \zeta(t) \zeta(t^{\prime}) \rangle =
2 C \delta(t - t^{\prime})$ ($C>0$). 
A possible demonstration of this relation is given in the Appendix. Notice that Eq. (\ref{langevin2}) is of the form of Eq. (\ref{langevin_general}) with $dW(t) = \zeta(t) dt$.

As we want to analyze the influences of the additive and multiplicative
noises separately, we will use the Eq. (\ref{langevin1}) instead of (\ref{langevin2}) henceforth.

Following the same procedure adopted to calculate the coefficients of the FPE associated with Eq. (\ref{langevin_general}), we can easily calculate the Kramers-Moyal coefficients for (\ref{langevin1}).
They are:
\begin{eqnarray}
K_\theta(u) &=& f(u) + 2 \theta M g(u) g^{\prime}(u) \\
D(u) &=& A + M \left[ g(u) \right]^2
\end{eqnarray}
These results lead us the following FPE:
\begin{widetext}
\begin{equation}
\frac{\partial P(u,t)}{\partial t} = -\frac{\partial}{\partial u} \left\{ \left[ f(u) + 2 \theta M g(u)
g^{\prime}(u) \right]P(u,t) \right\} + M \frac{\partial^2}{\partial u^2} \left\{ \left[g(u)\right]^2 P(u,t)
\right\} + A \frac{\partial^2}{\partial u^2} P(u,t) \; ,
\end{equation}
\end{widetext}
which is basically Eq. (\ref{FPE1}) with an additional term due to additive noise.

For $f(u)$ derived from a potential-like function $V(u) = (\tau/2)\left[g(u)\right]^2$, the stationary solution has the form $P(u,\infty) \propto e_q^{-\bar\beta V(u)}$ \cite{Anteneodo_Tsallis} where the $q$-exponential function is defined as follows: $e_q^x \equiv [1+(1-q)x]_+^{1/(1-q)}$, with $[z]_+ =z$ if $z>0$, and zero otherwise; $e_1^x=e^x$.
If we consider the simple case $g(u) \propto u$, then the stationary state distribution
is a $q$-Gaussian, i.e., $P(u,\infty) \propto e_q^{-\beta u^2}$, where $q$ and $\beta$ are given by:
\begin{eqnarray}
\beta &=& \frac{\tau + 2M (1 - \theta)}{2A} \\
\label{qtheta}q &=& \frac{\tau + 2M (2 - \theta)}{\tau + 2M (1 - \theta)}
\end{eqnarray}
We can verify that these results unify those obtained in \cite{Anteneodo_Tsallis}, more specifically $q= \frac{\tau + 4M }{\tau + 2M}$ for the It\^o case ($\theta = 0$), and $q= \frac{\tau + 3M}{\tau + M}$ for the Stratonovich case ($\theta=1/2$).

Let us remind at this point that the $q$-Gaussian form precisely is the one which, under appropriate constraints, extremizes the entropy
%
\begin{eqnarray}\label{Sq}
S_q[p] \equiv k_B \frac{1-\int dx \,[p(x)]^q}{q-1} \; ,
\end{eqnarray}
%
where $S_1 \equiv S_{BG} = - k_B \int dx p(x) \ln p(x)$.
For further connections between the structure of Fokker-Planck-like equations and entropy see \cite{evaldofernando,lisa}.

\section{Generalized moments and kurtosis} \label{escort}

As already mentioned, the $q$-Gaussian is the distribution form which extremizes the entropy (\ref{Sq}) under appropriate constraints. There are many theoretical reasons suggesting that, in the extremization of $S_q$ , it is convenient to express the constraints that are being imposed  in a escort mean value form. Escort mean values (or $q$-moments) are useful tools to analyze power law
distribution that frequently appear in the study of complex systems. They are defined as
\begin{equation}
 \left\langle A(x) \right\rangle_{q} = \int_{-\infty}^{+\infty} A(x)
f_{q}(x) dx,
\end{equation}
where the {\it escort probability density} is given by
\begin{equation}
 f_{q}(x) =
\frac{[f(x)]^{q}}{\int_{-\infty}^{+\infty}[f(x)]^{q} dx} \,.
\end{equation}

The characterization of a probability density in terms of its escort mean values
is a natural extension of the well-known characterization of a distribution in
terms of its standard moments if all of them are {\it finite}. The physical interpretation of $q$-generalized mean values demands an explanation. The quantities that are physically important in order to characterize a distribution are, for instance, the most probable value, the range of the typical values, the width of the distribution, its asymmetry, and so on. Such information is conveniently contained in the set of successive mean values of the distribution, {\it as long as they are finite}. What can be done whenever all moments above a given one diverge? For example, if we are dealing with the Cauchy-Lorentz distribution, how can we characterize its width (obviously {\it finite} for any given such distribution)? Certainly not through its second moment, since it diverges! By appropriately choosing the value of $q$ (see below), its width can be characterized by its $q$-variance (i.e., its variance with the escort distribution), which will also be {\it finite}.

In \cite{tsallis_plastino_alvarez}, the
correct set of all escort mean values is shown, together with the set of all associated
normalizing quantities, that characterize a given probability density $f(x)$, even if it decays as slowly as a power-law.
Based on a generalization of the Fourier transform, namely the $q$-Fourier
transform, defined by \cite{q-fourier}
\begin{equation}
\mathcal{F}_{q}[f](\xi) = \int_{-\infty}^{+\infty} f(x) \,
e_{q}^{i\xi[f(x)]^{q-1}} dx,
\qquad (q\geq1),
\end{equation}
it is possible to expand the $q$-characteristic function and obtain the
correct exponents values necessary to perform the calculations of the
$q$-moments. The family of escort mean values which arises from this procedure
are given by
\begin{equation}
 \left\langle x\right\rangle_{q_n}  = \frac{\int_{-\infty}^{+\infty} x^n[f(x)]^{q_n} \, dx}{\int_{-\infty}^{+\infty}[f(x)]^{q_n} dx} \,.
\end{equation}
where
\begin{equation}
 \label{Qgamma} q_n = 1 + n(q-1) \;.
\end{equation}

The kurtosis, usually defined in terms of the ratio between the fourth ($n=4$) and three
times the squared second moment ($n=2$) of a given distribution, is regarded as a
measure of how different a given distribution is from a Gaussian. Therefore the
$q$-kurtosis, defined in terms of $q$-moments, is the analogous measure for
$q$-Gaussians. Its mathematical forms are defined, respectively, as follows:
\begin{eqnarray}
\label{qkurtosis} \kappa_{q} = \frac{\int_{-\infty}^{+\infty} x^4
\left[ f(x)\right]^{4q-3} dx}{3 \left\lbrace \int_{-\infty}^{+\infty}
x^2 \left[ f(x)\right]^{2q-1} dx\right\rbrace ^2} \; .
\end{eqnarray}
Notice that the standard kurtosis can be regarded as a particular case ($q=1$) of Eq. (\ref{qkurtosis}). Notice also that the number 3 in the denominator of Eq. (\ref{qkurtosis}) plays no special role, and can equally well be replaced by the $q$-dependent value exactly corresponding to $q$-Gaussians, instead of that for Gaussians.

To calculate the correct value of $q$ we shall now consider a typical
situation arising in  complex systems, such as those commonly addressed
within $q$-thermostatistics. Usually these systems asymptotically behave
as power-law probability densities, i.e.,
\begin{equation}
 f(x)\sim|x|^{- \gamma}  \quad (|x|\rightarrow \infty; \gamma > 1) \; .
\end{equation}
It is clear that, if $f(x)$ is not defined on a bounded interval, the standard
linear moments $\langle x^n \rangle$ may diverge above some value of $n$.
Therefore the standard procedure to characterize the probability density through all its moments cannot be implemented. However, in \cite{tsallis_plastino_alvarez} it is shown how the escort mean values can overcome such difficulties.
In this work, the relation is established between the exponent $\gamma$ of the power-law distribution,
and the escort mean values parameter $q$. It is given by
\begin{equation}
 \label{Qgamma} q = 1 + \frac{1}{\gamma} \;.
\end{equation}
So, once we know the power-law exponent $\gamma$, we will have the
correct value of $q$, hence of  the values $q_n$ which enable the calculation
of the escort mean values of the probability density.

For the special case where $f(x) = e_Q^{-\beta x^2}$ (a $Q$-Gaussian), we have:
\begin{equation}
 e_{Q}^{-\beta x^2}\sim|x|^{-2/(Q-1)} \quad (|x|\rightarrow \infty) \, ,
\end{equation}
hence $\gamma = 2/(Q-1)$, for $Q>1$. In this case, the relation (\ref{Qgamma})
became:
\begin{equation}
 \label{qgamma} q-1 = \frac{Q-1}{2} \; .
\end{equation}
A $Q$-Gaussian is normalizable for $Q<3$; for $Q<1$ it has a compact support, and
an unbounded support for $1 \le Q<3$. Its second and fourth moments diverge for
$5/3<Q<3$ and $7/5<Q<3$, respectively. But their second and fourth $Q$-moments are finite
for $Q<3$. These two $Q$-moments can be analytically calculated
\cite{veit_nobre_tsallis,tsallis_plastino_alvarez} and written in terms of Gamma
functions,
\begin{eqnarray}
 \label{x2} \left\langle x^2 \right\rangle_{Q} &=& \left\lbrace \begin{array}{lcl}
	\frac{\beta^{-1} \varGamma \left( \frac{Q}{Q-1} -
\frac{3}{2}\right)}{2(Q-1)\varGamma \left( \frac{Q}{Q-1} - \frac{1}{2}\right)} & , &
\quad \mbox{for} \quad 1<Q<3\\
	\frac{\beta^{-1}}{2} & , & \quad \mbox{for} \quad Q=1\\
	\frac{\beta^{-1} \varGamma \left( \frac{Q}{1-Q} +
\frac{3}{2}\right)}{2(1-Q)\varGamma \left( \frac{Q}{1-Q} + \frac{5}{2}\right)} & , &
\quad \mbox{for} \quad 0<Q<1\\
	\end{array} \right. \\
\label{x4} \left\langle x^4 \right\rangle_{2Q-1} &=&  \left\lbrace
\begin{array}{lcl}
	\frac{3\beta^{-2} \varGamma \left( \frac{Q}{Q-1} -
\frac{3}{2}\right)}{4(Q-1)^2\varGamma \left( \frac{Q}{Q-1} +
\frac{1}{2}\right)} & , & \quad \mbox{for} \quad 1<Q<3\\
	\frac{3\beta^{-1}}{4} & , & \quad \mbox{for} \quad Q=1\\
	\frac{3\beta^{-2} \varGamma \left( \frac{Q}{1-Q} +
\frac{1}{2}\right)}{4(1-Q)^2\varGamma \left( \frac{Q}{1-Q} +
\frac{5}{2}\right)} & , & \quad \mbox{for} \quad 0<Q<1 \;.\\
	\end{array} \right. 
\end{eqnarray}

\begin{figure}
 \centering
 \includegraphics[scale=0.29]{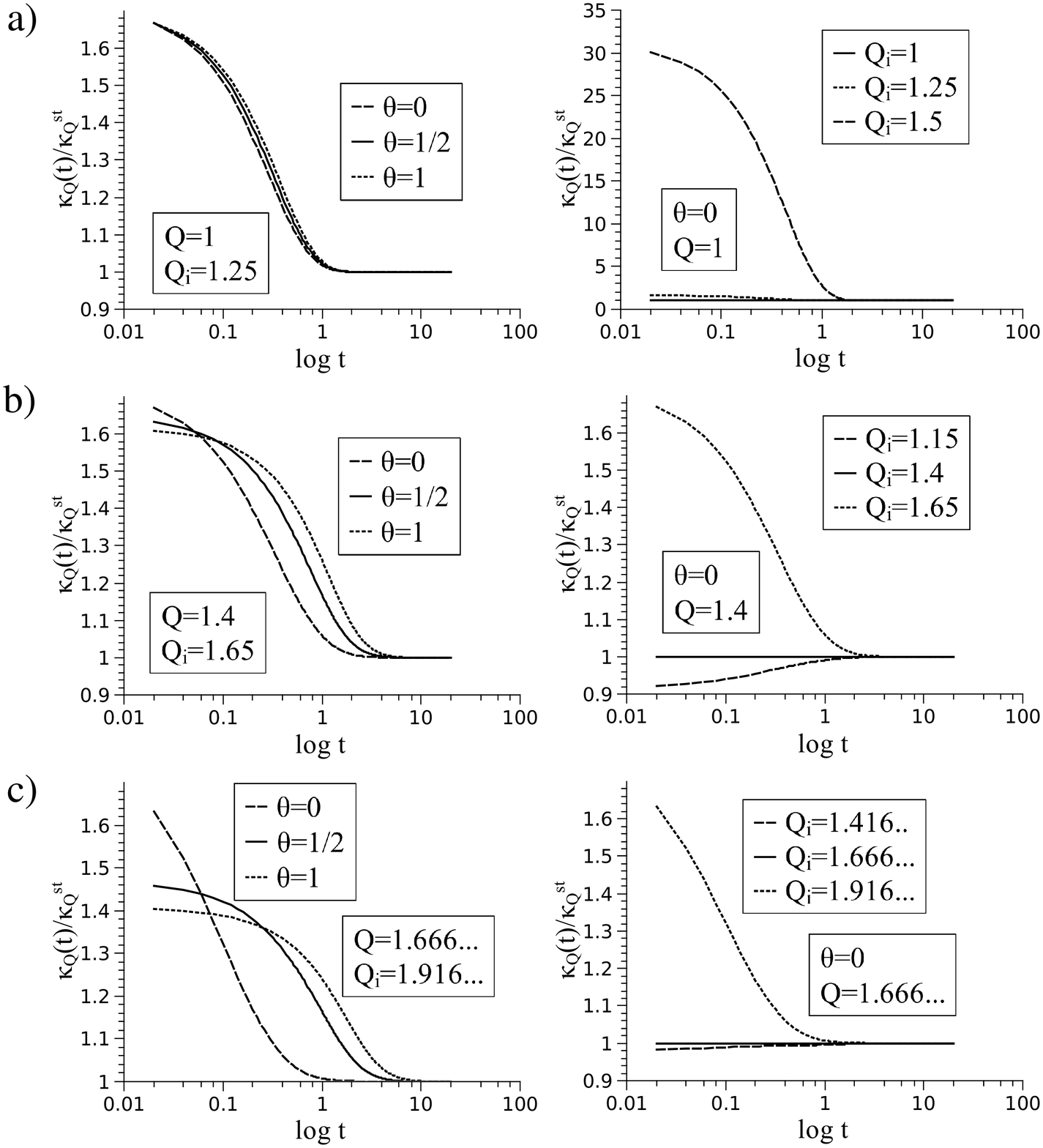}
 \caption{On the left column, the temporal evolution of $Q$-kurtosis for It\^o ($\theta=0$), Stratonovich 
($\theta=1/2$) and backward It\^o ($\theta=1$) stochastic procedures. The numerical results are
normalized by its stationary value $\kappa_Q^{st}=\kappa_Q(\infty)$. In (a) we have $Q=1$ and $Q_i=5/4$, in (b) $Q=7/5$ and
$Q_i=1.65$ and (c) $Q=1.666...$ and $Q_i=1.91666...$ .
As expected, for sufficient large times the numerical results approaches to unity. The right column shows the temporal evolution of the normalized $Q$-kurtosis with a fixed $\theta$. In (a) we have $Q=1$ and $Q_i=1,\, 5/4\, \text{and}\, 3/2$; (b) shows the results for $Q=7/5$ and $Q_i=1.15,\, 7/5\, \text{and}\, 1.65$; and (c) the results for $Q=5/3$ and $Q_i=1.41666...,\, 5/3\, \text{and}\, 1.91666...$\,.}
 \label{fig:1}
\end{figure}

We intend to compare the temporal evolution of the distributions given by the
FPE of the previous section to its stationary states, which we already know to be $Q$-Gaussians.
To do it, it is convenient to normalize the $Q$-kurtosis by its stationary value,
$\kappa_Q(\infty)=\lim_{t\rightarrow\infty}\kappa_Q(t)$, which is
given by Eqs. (\ref{x2}) and (\ref{x4}),
\begin{equation}
\label{kq} \kappa_Q(\infty) = \left\lbrace \begin{array}{lcl}
	\frac{\varGamma^2 \left( \frac{Q}{Q-1} - \frac{1}{2}\right)}{\varGamma
\left( \frac{Q}{Q-1} + \frac{1}{2}\right) \varGamma \left( \frac{Q}{Q-1} -
\frac{3}{2}\right)} & , & \quad \mbox{for} \; 1<Q<3\\
	1 & , & \quad \mbox{for} \quad Q=1\\
	\frac{\varGamma^2 \left( \frac{Q}{1-Q} + \frac{1}{2}\right) \varGamma
\left( \frac{Q}{1-Q} + \frac{5}{2}\right)}{\varGamma^2 \left( \frac{Q}{1-Q} +
\frac{3}{2}\right)} & , & \quad \mbox{for} \; 0<Q<1 \;.\\
	\end{array} \right. 
\end{equation} 
In Fig. \ref{fig:1} we can see the temporal evolution of the normalized
$Q$-kurtosis
obtained by the numerical integration of the FPE. In the left column, we show
the temporal
evolution for the three stochastic procedures above mentioned, It\^o
($\theta=0$), Stratonovich 
($\theta=1/2$) and backward It\^o ($\theta=1$). In the right column we
show the same
temporal evolution of the normalized $Q$-kurtosis but now with $\theta=0$ and
different values of $Q_i$.
As expected, the kurtosis approaches monotonically its stationary value
 for all values of parameter $Q_i$.

\begin{figure}
 \centering
 \includegraphics[scale=0.46]{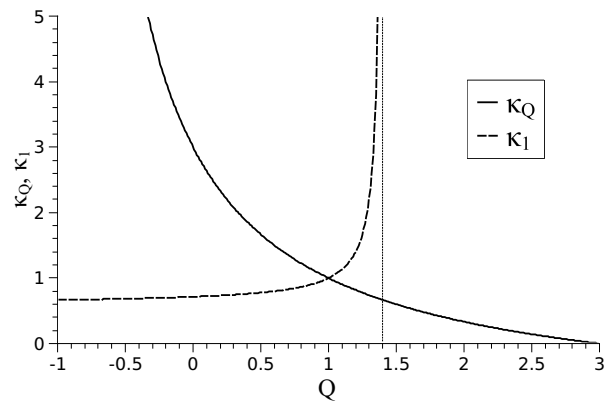}
 \caption{The standard kurtosis $\kappa_1$ and the $Q$-kurtosis $\kappa_Q$ as functions of $Q$. We verify
that both have the same value 1 for $Q=1$. For $Q>1$, the use of $\kappa_Q$ is generically more convenient than using
$\kappa_1$. Indeed, the latter one diverges at $Q=7/5$ 
while $\kappa_Q$ remains finite up to the maximal value $Q=3$. In contrast, for $Q<1$, $\kappa_1$ is more convenient than $\kappa_Q$. Indeed, $\kappa_Q$ diverges at $Q=-1$, whereas $\kappa_1$ remains finite down to $Q\to - \infty$. In the $Q \to - \infty$ limit, $\kappa_1$ saturates at the value $3/5$. }
 \label{fig:2}
\end{figure}

In Fig. \ref{fig:2} we can compare the behavior of the standard kurtosis and the $Q$-kurtosis
given by Eq. (\ref{kq}). As we can see, the $Q$-kurtosis diverges at
$Q=-1$, but is finite for $-1<Q<3$, while the standard kurtosis diverges for $7/5<Q<3$.
As expected, the $Q$-kurtosis coincides with the standard one for $Q=1$. Furthermore, the $Q$-kurtosis is a monotonically
decreasing function of $Q$. Based on this feature, experimentalists could use
it as a calibration curve that allows the determination of the proper value of the
entropic index $Q$ for a given system.

\section{Conclusions}

Based on stochastic integrals which unify It\^o and Stratonovich approaches, we obtain (in two different though equivalent forms) an unified Fokker-Planck equation which recovers, as particular instances, equations currently available in the literature. Both Fokker-Planck forms exhibit an explicit dependence on the unifying parameter ($\theta$). One of these forms shows this dependence only in the deterministic part as a noise-induced drift term; in the other one, it only appears in the diffusion term.
We present, based on escort mean values, an explicit form for the $Q$-generalized kurtosis to study the convergence towards the $Q$-Gaussian stationary solution. The difference in the convergence for the same $q$-Gaussian with different $\theta$'s is due to the change in the amplitude of the multiplicative noise needed to compensate the change in $\theta$ (see Eq. (\ref{qtheta})). This is the origin of the difference in the convergence velocity when we keep $Q_i$'s and $Q$'s constant, and vary $\theta$. Looking at the time evolution of the $Q$-kurtosis starting from different initial conditions, i.e. different $Q_i$'s, we see that the convergence is reached for a little less than two decades in all explored cases. Finally, we propose that, if a system is well described by the nonextensive statistical mechanics, and has a $Q$-Gaussian form for its probability density function, the $Q$-kurtosis can be used as a calibration curve to determine, from data, the best value of the entropic index $Q$.

\section*{Acknowledgments}

We acknowledge partial financial support from the Brazilian agencies
Conselho Nacional de Desenvolvimanto Cient\'ifico e Tecnol\'ogico (CNPq) and
Funda\c{c}\~ao Carlos Chagas Filho de Amparo \`a Pesquisa do Estado do Rio de Janeiro
(Faperj).

\section*{\appendixname}

\subsection{The integral $ I_\theta[W(t)] = \int_{t_0}^{t} dW(t^{\prime}) W(t^{\prime})$}\label{apendice1}

Here we perform, in details, the calculation of the stochastic integral of the
function $G(t) = W(t)$ in terms of discrete sums.

The definition of the unified stochastic integral is
\begin{widetext}
\begin{equation}
I_{\theta}[G(t)] \equiv \text{ms -}\hspace{-0.1cm}\lim_{n \rightarrow \infty} S_n =
\text{ms -}\hspace{-0.1cm}\lim_{n \rightarrow \infty} \sum_{i=1}^{n}
\left[\theta G_{i} + (1-\theta) G_{i-1}\right] \Delta W_i \; .
\end{equation}
\end{widetext}
If $G(t) = W(t)$, we have
\begin{equation}
S_n = \sum_{i=1}^{n} \left[\theta W_{i} \Delta W_{i} + (1-\theta) W_{i-1} \Delta W_{i} \right] \; .
\end{equation}
The terms $W_{i} \Delta W_{i}$ and $W_{i-1} \Delta W_{i}$ can be written as
\begin{eqnarray}
2 W_{i} \Delta W_{i} = (W_{i})^2 + (\Delta W_i)^2 - (\overbrace{W_{i} - \Delta W_i}^{W_{i-1}})^2 &&\\
2 W_{i-1} \Delta W_{i} = (\underbrace{W_{i-1} + \Delta W_i}_{W_{i}})^2 - (W_{i-1})^2 - (\Delta W_i)^2 &&\; .
\end{eqnarray}
Notice that in both equations we have terms like $(W^2_i - W^2_{i-1})$ under
a summation. Therefore only the first and the last terms need to be taken into
account in the summation. Indeed, all the others mutually cancel, and we will finally have 
\begin{eqnarray}
\sum_{i=1}^{n} W_{i} \Delta W_{i} = \frac{W^2(t) - W^2(t_0) + \sum_{i=1}^{n}
(\Delta W_i)^2}{2} && \\
\sum_{i=1}^{n} W_{i-1} \Delta W_{i} = \frac{W^2(t) - W^2(t_0) - \sum_{i=1}^{n}
(\Delta W_i)^2}{2} && \; .
\end{eqnarray}
Taking this last two equations into account and the fact that
$\text{ms -}\lim_{n \rightarrow \infty} \sum_{i=1}^{n} (\Delta W_{i})^2 = t - t_0$,
we obtain
\begin{widetext}
\begin{eqnarray}
I_{\theta}[W(t)] \equiv \text{ms -}\hspace{-0.1cm}\lim_{n \rightarrow \infty} S_n = \frac{1}{2} \left\{ [W(t)]^2  - [W(t_0)]^2 - (1 - 2\theta) (t - t_0) \right\}\; .
\end{eqnarray}
\end{widetext}
which is the same result presented in Eq. (\ref{example}).

\subsection{How to get Eq. (\ref{FPE1}) from Eq. (\ref{FPE2})}\label{apendice2}

In section \ref{FPE} we claim that Eqs. (\ref{FPE1}) and (\ref{FPE2}) are equivalent. To prove this statement, we only need to show the following equality: 
\begin{equation}
\nonumber K_0 P - \frac{1}{2}[D]^{2\theta}\frac{\partial}{\partial u} \left\lbrace [D]^{1-2\theta} P\right\rbrace = K_\theta P - \frac{1}{2}\frac{\partial}{\partial u} \left[D P \right] \; ,
\end{equation}
where the dependence on the dynamical variable $u$ was omitted.
To do this, we perform the derivative on the left hand side.
%
\begin{eqnarray}
\nonumber D^{2\theta}\frac{\partial}{\partial u} \left[  D^{1-2\theta} P \right]  &=& D^{2\theta} \left[ (1-2 \theta) D^{-2\theta} P \frac{\partial D}{\partial u} + D^{1-2\theta} \frac{\partial P}{\partial u}\right] \\
\nonumber &=& -2\theta \frac{\partial D}{\partial u} P + \left[P \frac{\partial D}{\partial u} + D \frac{\partial P}{\partial u} \right] \\
&=& -2\theta \frac{\partial D}{\partial u} P + \frac{\partial}{\partial u} \left[ DP \right] \; .
\end{eqnarray}
%
So,
\begin{widetext}
\begin{eqnarray}
\nonumber K_0 P - \frac{1}{2}[D]^{2\theta}\frac{\partial}{\partial u} \left\lbrace [D]^{1-2\theta} P\right\rbrace &=& K_0 P - \frac{1}{2} \left\lbrace -2\theta \frac{\partial D}{\partial u} P + \frac{\partial}{\partial u} \left[ DP \right]  \right\rbrace \\
&=& K_0 P + \theta \frac{\partial D}{\partial u} P - \frac{1}{2} \frac{\partial}{\partial u} \left[ DP \right] \; ,
\end{eqnarray}
\end{widetext}
where $K_0 P + \theta \partial_u D P = K_{\theta}$.

\subsection{The equivalence between $\dot{u} = f(u) + g(u)\xi(t) + \eta(t)$ and $\dot{u} = f(u) + \tilde{g}(u)\zeta(t)$}\label{apendice3}

Suppose two different Langevin equations for the stochastic processes $u(t)$,
both with the same 
deterministic drift force $f(u)$ but different noise sources,
\begin{eqnarray}
\dot{u} &=& f(u) + g(u)\xi(t) + \eta(t) \\
\dot{u} &=& f(u) + \tilde{g}(u)\zeta(t) \; .
\end{eqnarray}
The mean-value solution of both equations is
\begin{equation}
\langle u(t) \rangle = \int_0^t dt^{\prime} f(u) \; .
\end{equation}
Let us calculate the variance corresponding to both equations. From the first one we obtain
\begin{widetext}
\begin{eqnarray}\label{ap1}
\langle [u(t) - \langle u(t) \rangle ]^2 \rangle &=& \left\langle \int_0^t
dt^{\prime} \int_0^t dt^{\prime\prime} \left\{ g[u(t^{\prime})] \xi(t^{\prime})
+ \eta(t^{\prime}) \right\} \left\{ g[u(t^{\prime\prime})] \xi(t^{\prime\prime})
+ \eta(t^{\prime\prime}) \right\} \right\rangle \\
&=& \int_0^t dt^{\prime} \int_0^t dt^{\prime\prime} \left\{g[u(t^{\prime})]
g[u(t^{\prime\prime})] \underbrace{\langle \xi(t^{\prime}) \xi(t^{\prime\prime})
\rangle}_{=2M\delta(t^{\prime} - t^{\prime\prime})} + \underbrace{\langle
\eta(t^{\prime}) \eta(t^{\prime\prime}) \rangle}_{=2A\delta(t^{\prime} -
t^{\prime\prime})} \right. \\
&& \qquad \qquad \qquad \left. + g[u(t^{\prime})] \underbrace{\langle
\xi(t^{\prime}) \eta(t^{\prime\prime}) \rangle}_{=0} + g[u(t^{\prime\prime})]
\underbrace{\langle \xi(t^{\prime\prime}) \eta(t^{\prime}) \rangle}_{=0}
\right\} \\
&=& \int_0^t dt^{\prime} \left\lbrace 2M \left\{ g[u(t^{\prime})] \right\}^2 +
2A \right\rbrace \; .
\end{eqnarray}
\end{widetext}
From the second one we obtain
\begin{widetext}
\begin{eqnarray}\label{ap2}
\langle [u(t) - \langle u(t) \rangle ]^2 \rangle &=& \int_0^t dt^{\prime}
\int_0^t dt^{\prime\prime} \tilde{g}[u(t^{\prime})]
\tilde{g}[u(t^{\prime\prime})] \underbrace{\langle
\zeta(t^{\prime})\zeta(t^{\prime\prime}) \rangle}_{2C\delta(t^{\prime} -
t^{\prime\prime})} \\
&=& \int_0^t dt^{\prime} 2C \left\{ \tilde{g}[u(t^{\prime})] \right\}^2 \; .
\end{eqnarray}
\end{widetext}
If we want that both Langevin equations give us the same mean-squared value for
the stochastic 
process $u(t)$, we must match the integrands of Eqs. (\ref{ap1}) and
(\ref{ap2}). This leads to the following relation among noise terms:
\begin{equation}\label{eff_noise}
\tilde{g}(u) = \sqrt{\frac{M [g(u)]^2 + A}{C}} \; .
\end{equation}
This result shows that Eqs. like (\ref{langevin1}) can be rewritten as in
(\ref{langevin2}), with 
an effective multiplicative noise term given by Eq. (\ref{eff_noise}).



\begin{thebibliography}{99}

\bibitem{Gardiner} C. W. Gardiner, {\it Handbook of Stochastic Methods} (Springer, Berlin, 1997).

\bibitem{Risken} H. Risken, \textit{The Fokker-Planck Equation. Methods of
solutions and applications} (Springer-Verlag, New York, 1984).

\bibitem{Germano}G. Germano, M. Politi, E. Scalas and R. L. Schilling, 
Phys. Rev. E \textbf{79}, 066102 (2009).

\bibitem{Lau} A. W. C. Lau and T. C. Lubensky,
 Phys. Rev. E \textbf{76}, 011123 (2007)

\bibitem{Lancon1} P. Lan\c{c}on, G. Batrouni, L. Lobry and N. Ostrowsky, 
Physica A \textbf{304}, 65 (2002)

\bibitem{Lancon2} P. Lan\c{c}on, G. Batrouni, L. Lobry and, N. Ostrowsky, 
Euro-Phys. Lett. \textbf{54}, 28 (2001).

\bibitem{backward} L. Chung and Jean-Claude Zambrini,
Monograph of the Portuguese Mathematical Society, Vol. 1, World Scientific Publishing Company, 2003;
A. I. Ilin, R. Z. Khasminski, Theo. Prob. Appl. \textbf{9}, 466 (1964);
E. Nelson, {\it Dynamical Theories of the Brownian Motion}, (Princeton, University Press, Princeton, 1967);
E. Nelson, {\it Quantum Fluctuations}, (Princeton University, Princeton, 1985.)

\bibitem{causality} N. Dokuchaev,
arXiv:0708.2497v1.

\bibitem{Anteneodo_Tsallis} C. Anteneodo and C. Tsallis,
J. Math. Phys. \textbf{44}, 5194, (2003).

\bibitem{Kampen} N. van Kampen, \textit{Stochastic Processes in Physics and Chemistry}
(North-Holland, Amsterdam, 1981).

\bibitem{van_KampenJSP} N. van Kampen, 
J. of Stat. Phys. \textbf{24}, 175, (1981).

\bibitem{evaldofernando}V. Schwammle, E.M.F. Curado and F.D. Nobre, 
Eur. Phys. J. B {\bf 58}, 159 (2007); V. Schwammle, F.D. Nobre and E.M.F. Curado, 
Phys. Rev. E {\bf 76}, 041123 (2007).

\bibitem{lisa} L. Borland, F. Pennini, A. R. Plastino and A. Plastino, 
Eur. Phys. J. B {\bf 12}, 285 (1999).

\bibitem{Beckbook}C. Beck and F. Schlogl, {\it Thermodynamics of Chaotic Systems} (Cambridge University Press, Cambridge, 1993).

\bibitem{Tsallis1998} C. Tsallis, R. S. Mendes and A. R. Plastino,
Physica A \textbf{261} 534 (1998).

\bibitem{Tsallis2009}C. Tsallis, 
J. Stat. Phys. {\bf 52}, 479 (1988); C. Tsallis, {\it Introduction to Nonextensive Statistical Mechanics - Approaching a Complex World} (Springer, New York, 2009).

\bibitem{UpadhyayaRieuGlazierSawada2001}A. Upadhyaya, J.-P. Rieu, J.A. Glazier and Y. Sawada, 
Physica A {\bf 293}, 549 (2001).

\bibitem{Reynolds2010}A.M. Reynolds, 
Physica A {\bf 389}, 273-277 (2010).

\bibitem{DanielsBeckBodenschatz2004}K.E. Daniels, C. Beck and E. Bodenschatz, 
Physica D {\bf 193}, 208 (2004).

\bibitem{ArevaloGarcimartinMaza2007a}R. Arevalo, A. Garcimartin and D. Maza, 
Eur. Phys. J. E {\bf 23}, 191-198 (2007)[DOI10.1140/epje/i2006-10174-1].

\bibitem{ArevaloGarcimartinMaza2007b}R. Arevalo, A. Garcimartin and D. Maza, 
{\it A non-standard statistical approach to the silo discharge}, in {\it Complex Systems - New Trends and Expectations}, eds. H.S. Wio, M.A. Rodriguez and L. Pesquera, Eur. Phys. J.-Special Topics {\bf 143} (2007) [DOI: 10.1140/epjst/e2007-00087-9].

\bibitem{DouglasBergaminiRenzoni2006}P. Douglas, S. Bergamini and F. Renzoni, 
Phys. Rev. Lett. {\bf 96}, 110601 (2006); G.B. Bagci and U. Tirnakli, 
Chaos {\bf 19}, 033113 (2009).

\bibitem{LiuGoree2008}B. Liu and J. Goree, 
Phys. Rev. Lett. {\bf 100}, 055003 (2008).

\bibitem{DeVoe2009}R.G. DeVoe, 
Phys. Rev. Lett. {\bf 102}, 063001 (2009).

\bibitem{Borland2002a}L. Borland, 
Phys. Rev. Lett. {\bf 89}, 098701 (2002).

\bibitem{Borland2002b}L. Borland, 
Quantitative Finance {\bf 2}, 415 (2002).

\bibitem{Queiros2005}S.M.D. Queiros, 
Quant. Finance {\bf 5}, 475-487 (2005).

\bibitem{BurlagaVinas2005}L.F. Burlaga and A.F.-Vinas, 
Physica A {\bf 356}, 375 (2005).

\bibitem{BurlagaNess2009}L.F. Burlaga and N.F.  Ness, 
Astrophys. J.  {\bf 703}, 311-324   (2009). 

\bibitem{BakarTirnakli2009}B. Bakar and U. Tirnakli, 
Phys. Rev. E {\bf 79}, 040103(R) (2009).

\bibitem{CarusoPluchinoLatoraVinciguerraRapisarda2007}F. Caruso, A. Pluchino, V. Latora, S. Vinciguerra and A. Rapisarda,
Phys. Rev. E {\bf 75}, 055101(R)(2007). 

\bibitem{MoyanoAnteneodo2006}L.G. Moyano and C. Anteneodo, 
Phys. Rev. E {\bf 74}, 021118 (2006).

\bibitem{CarvalhoSilvaNascimentoMedeiros2008}J.C. Carvalho, R. Silva, J.D. do Nascimento and J.R. de Medeiros, 
Europhys. Lett. {\bf 84}, 59001 (2008). 

\bibitem{CMS2010}CMS Collaboration, 
J. High Energy Phys. {\bf 02}, 041 (2010).

\bibitem{PickupCywinskiPappasFaragoFouquet2009}R.M. Pickup, R. Cywinski, C. Pappas, B. Farago and P. Fouquet, 
Phys. Rev. Lett. {\bf 102}, 097202 (2009).

\bibitem{veit_nobre_tsallis} V. Schwammle, F. D. Nobre and C. Tsallis,
Eur. Phys. J. B \textbf{66}, 537
(2008).

\bibitem{tsallis_plastino_alvarez} C. Tsallis, A. R. Plastino and
R. F. Alvarez-Estrada, 
J. Math. Phys. \textbf{50}, 043303 (2009).

\bibitem{q-fourier} S. Umarov, C. Tsallis an S. Steiberg, Milan J. of
Mathematics \textbf{76}, 307 (2008); S. Umarov, C. Tsallis, M. Gell-Mann and S. Steinberg,  
J. Math. Phys. {\bf 51}, 033502 (2010).

\end{thebibliography}
\end{document}